\documentclass{aastex62}
\usepackage[utf8]{inputenc}
\usepackage{natbib}
\setcitestyle{square, numbers, naturemag}
\usepackage{graphicx}

\renewcommand{\apjl}{Astrophys. J. Lett.}
\renewcommand{\apjs}{Astrophys. J. Suppl.}
\renewcommand{\aap}{Astron. Astrophys.}

\begin{document}
\title{Highlights of Exoplanetary Science from {\it Spitzer}}

\author{Drake Deming} 
\affiliation{Department of Astronomy, University of Maryland, College Park, MD 20742, USA; ddeming@astro.umd.edu}
\affiliation{NASA Astrobiology Institute's Virtual Planetary Laboratory}
\author{Heather Knutson}
\affiliation{Division of Geological and Planetary Sciences, California Institute of Technology, Pasadena, CA 91125, USA; hknutson@caltech.edu} 

\begin{abstract}
Observations of extrasolar planets were not projected to be a significant part of the \emph{Spitzer Space Telescope's} mission when it was conceived and designed.  Nevertheless, \emph{Spitzer} was the first facility to detect thermal emission from a hot Jupiter, and the range of \emph{Spitzer's} exoplanetary investigations grew to encompass transiting planets, microlensing, brown dwarfs, and direct imaging searches and astrometry.  \emph{Spitzer} used phase curves to measure the longitudinal distribution of heat as well as time-dependent heating on hot Jupiters.  \emph{Spitzer's} secondary eclipse observations strongly constrained the dayside thermal emission spectra and corresponding atmospheric compositions of hot Jupiters, and the timings of eclipses were used for studies of orbital dynamics. \emph{Spitzer's} sensitivity to carbon-based molecules such as methane and carbon monoxide was key to atmospheric composition studies of transiting exoplanets as well as imaging spectroscopy of brown dwarfs, and complemented Hubble spectroscopy at shorter wavelengths.  \emph{Spitzer's} capability for long continuous observing sequences enabled searches for new transiting planets around cool stars, and helped to define the architectures of planetary systems like TRAPPIST-1. \emph{Spitzer} measured masses for small planets at large orbital distances using microlensing parallax.  {\it Spitzer} observations of brown dwarfs probed their temperatures, masses, and weather patterns.  Imaging and astrometry from {\it Spitzer} was used to discover new planetary mass brown dwarfs and to measure distances and space densities of many others.
\\
\\
\end{abstract}

\section{Introduction} \label{sec:intro}
The first detection of a Jupiter-sized planet orbiting a solar-type star with a 4-day period \citep{mayor05} carried significant implications for the characterization of exoplanets.  It was immediately clear that such planets were likely to be quite hot ($T > 1000$K) due to strong stellar irradiation, and their rotation was likely to be tidally locked to their short period orbits \citep{guillot96}. The combination of large size and high temperature made these ``hot Jupiters" very amenable to observations in the infrared (IR) \citep{seagersasselov00, seagerwhitney00}. Indeed, observations of exoplanets - both hot and cold, and both large and small - was a major science theme for {\it Spitzer}. 

In this paper, we review some highlights of {\it Spitzer's} exoplanetary science, starting with the temperature structure and chemistry of hot transiting exoplanets (Section~\ref{sec:temp_chem}), and continuing with the implications for their atmospheric dynamics (Section~\ref{sec:atm_dynamics}), and the properties of their orbits (Section~\ref{sec:orb_dynamics}). Although the majority of exoplanets studied by {\it Spitzer} were in transiting systems, {\it Spitzer} investigators also used other techniques such as imaging (Section~\ref{subsec:imaging}) and microlensing (Section~\ref{subsec:microlensing}).  Given that exoplanets overlap the mass range of brown dwarfs, we review highlights of {\it Spitzer's} brown dwarf science in Section~\ref{sec:brown_dwarfs}.  We comment on how {\it Spitzer} set the stage for the James Webb Space Telescope in Section~\ref{sec:stage}. For a previous review of exoplanetary science using {\it Spitzer}, see \citet{beichman18}. Protoplanetary and debris disks are closely related to exoplanets, and disk science from {\it Spitzer} is reviewed by \citet{chen20} in this review series.

\begin{figure*}[h!]
\centering
\includegraphics[width=4in]{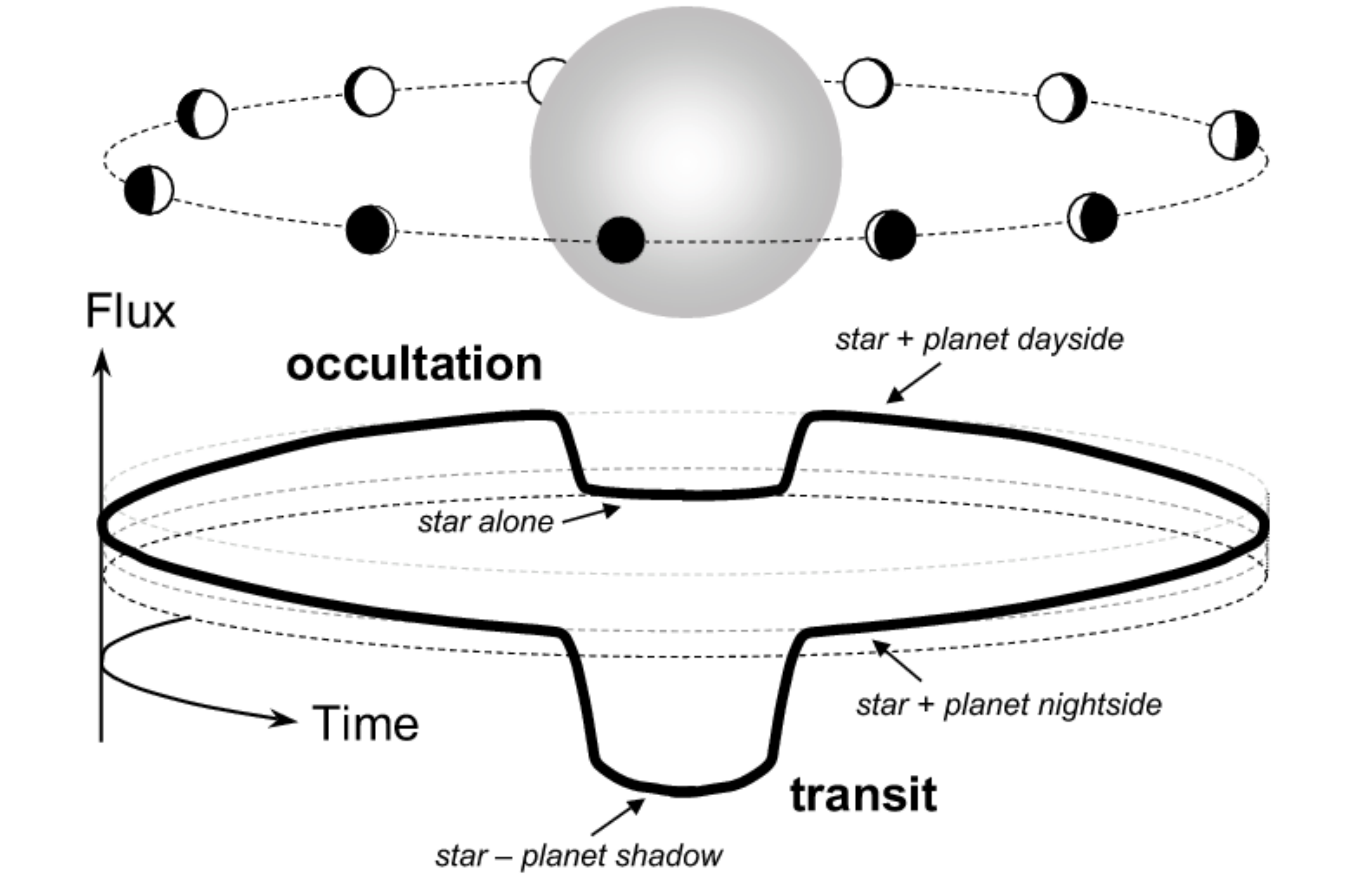}
\caption{Illustration of the techniques used by {\it Spitzer} to measure transiting planets, from \cite{winn10}.  The occultation is an alternate term for secondary eclipse, and it allows the emergent radiation from the planet to be separated from the star.  The depth of the primary transit will vary with wavelength, because molecular and atomic absorptions make the atmosphere effectively more extended at some wavelengths.  The brightness of the system as a function of orbital phase is called the phase curve.  Note that {\it Spitzer} did not spatially resolve any of this structure; the measurements were made using time variations in the total light of the system.}
\label{fig:techniques}
\end{figure*}

\begin{figure*}[h!]
\centering
\includegraphics[width=6in]{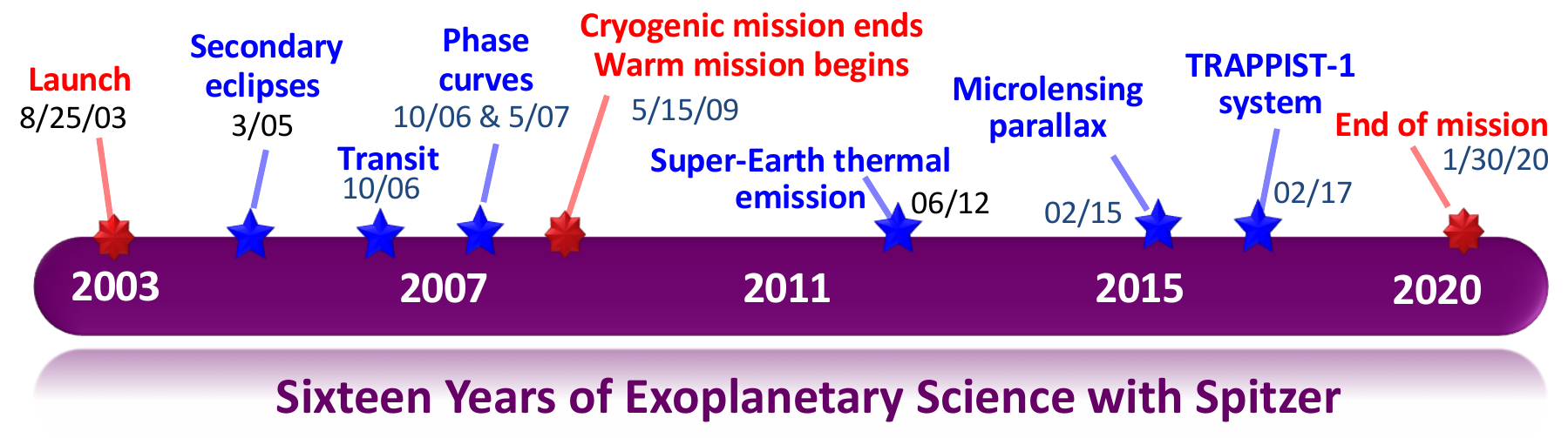}
\caption{Timeline of some major exoplanetary scientific highlights from {\it Spitzer}. Programmatic milestones are illustrated in red, and scientific highlights in blue.  {\it Spitzer's} pioneering scientific highlights include detection of hot Jupiter thermal emission via secondary eclipses \citep{charbonneau05, deming05}, measurement of phase curves \citep{harrington06, knutson07}, the detection of thermal emission from a super-Earth \citep{demory12}, measurement of microlensing parallax \citep{udalski15}, and defining the architecture of the TRAPPIST-1 system (seven nearly co-planar planets) \citep{gillon17c}.}
\label{fig:timeline}
\end{figure*}

The major techniques that {\it Spitzer} used for transiting planets (transits, secondary eclipses, and phase curves) are illustrated in Figure~\ref{fig:techniques}. Figure~\ref{fig:timeline} shows a summary timeline, illustrating when some of the important exoplanet milestones occurred during the mission.
\clearpage

\section{Temperature Structure and Chemistry}\label{sec:temp_chem}

Prior to the launch of {\it Spitzer}, the atmospheric chemistry of hot Jupiters was projected to be dominated by water vapor, carbon monoxide, and also (depending on the temperature), methane \citep{burrows97, burrowssharp99, seagerwhitney00, seagersasselov00}.  The hottest of these planets were expected to have low albedos because they would be too hot for cloud condensation \citep{sudarsky00}.  {\it Spitzer} confirmed those basic expectations, but variations on the models have been found, as we describe below.

\subsection{The First Detections of Dayside Emission Spectra}\label{subsec:emergent}

{\it Spitzer} was the first telescope to detect the infrared radiation emitted by transiting exoplanets using the secondary eclipse technique.  Subtracting spectra or photometry taken during eclipse (planet behind star) from measurements outside of eclipse (star + planet contributing) yielded the emergent spectrum of the exoplanet's dayside atmosphere.  The first secondary eclipse detections \citep{charbonneau05, deming05} were quickly followed by theoretical interpretations \citep{barman05, burrows05, seager05} and by additional secondary eclipse measurements \citep{deming06}.  The earliest measurements focused on hot Jupiters transiting bright stars ($V_{mag}<8$), such as HD\,189733 \citep{bouchy05} and HD\,209458 \citep{charbonneau00}.  As additional transiting hot Jupiters were discovered by ground-based transit surveys, {\it Spitzer} observers used secondary eclipses to construct broadband emission spectra for the dayside atmospheres of these planets.  

\begin{figure*}[h]
\centering
\includegraphics[width=3.4in]{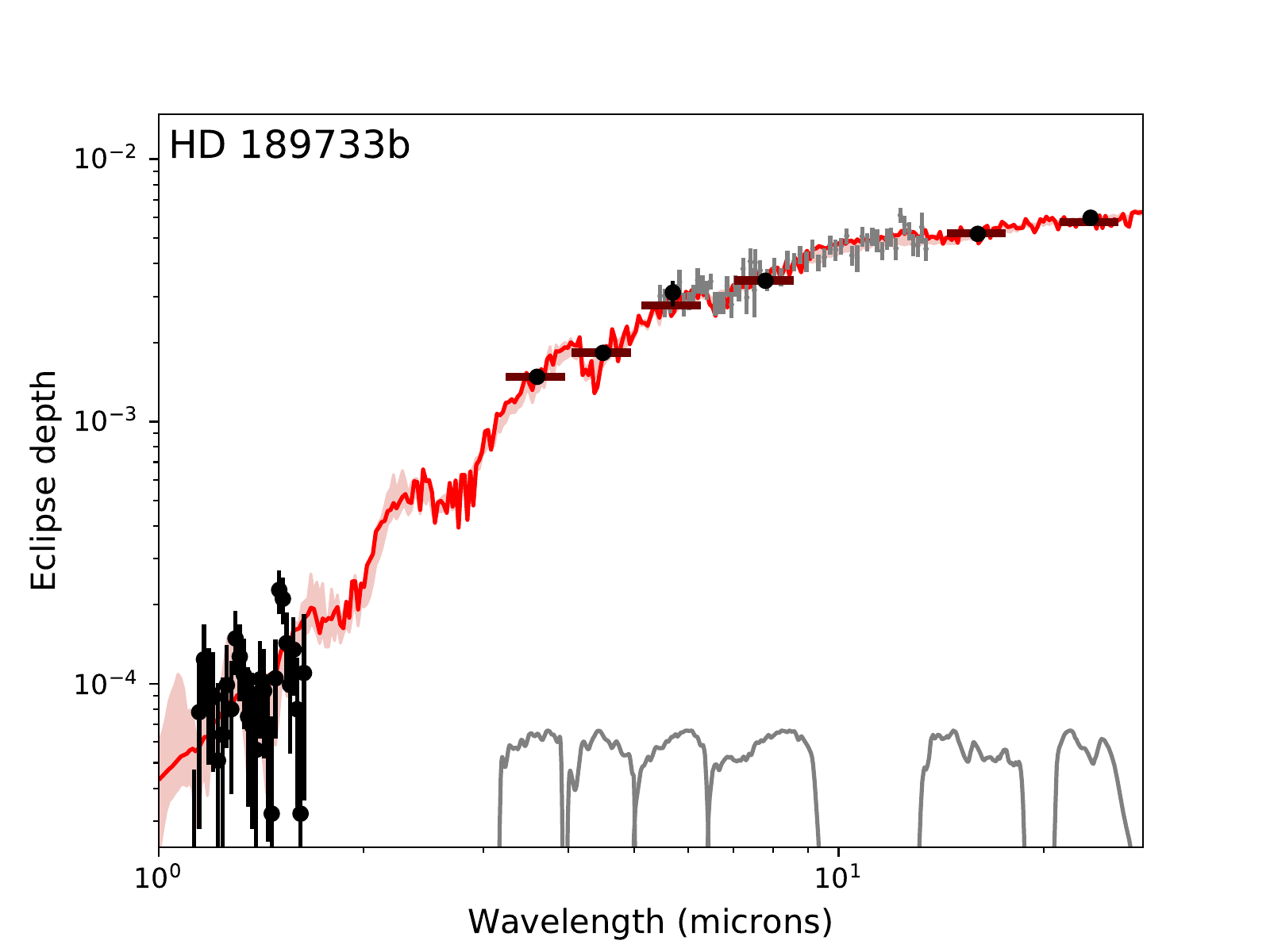}
\includegraphics[width=3.4in]{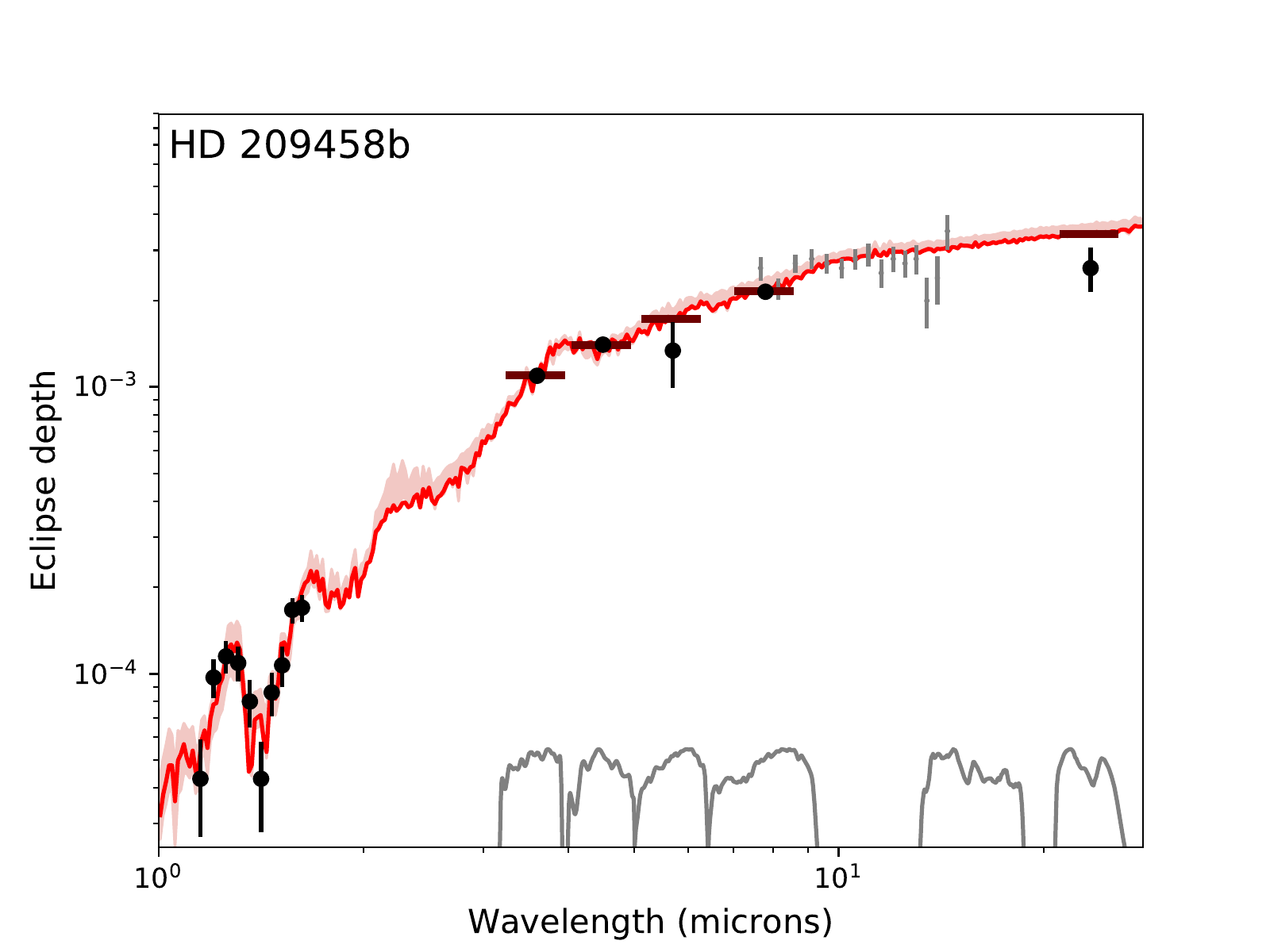}
\caption{\emph{HST} ($1-2$\,$\mu$m) and {\it Spitzer} ($3-24$\,$\mu$m) dayside thermal emission spectra of the two brightest hot Jupiters: HD\,189733b (left), and HD\,209458b (right). For HD\,189733b, black points are \emph{HST} spectroscopy \citep{crouzet14}, and \emph{Spitzer} photometry \citep{knutson12}; grey points are \emph{Spitzer} spectroscopy \citep{todorov14}. For HD\,209458b black points are \emph{HST} spectroscopy \citep{line16}, and \emph{Spitzer} photometry \citep{deming05, diamond-lowe14, kilpatrick19}; grey points are \emph{Spitzer} spectroscopy, \citep{richardson07, swain08}. Red lines are best-fit models from the open-source atmosphere retrieval code PLATON (\citealp{zhang19} and Zhang et al., ApJ submitted), with model uncertainties as light red shading.  \emph{Spitzer}/IRS spectroscopic data \citep{todorov14} are plotted, but were not used in the fit: they are offset by +900 ppm (HD\,189733b) and -500 ppm (HD\,209458) to match the best-fit PLATON model.  The models were integrated over each of the \emph{Spitzer} photometric bands (band transmission shown in grey), with the band-integrated model values plotted as dark red horizontal lines.  Plots are courtesy of M. Zhang.}
\label{fig:189_spectrum}
\end{figure*}

Those observations were in basic accord with early theoretical models \citep{burrows97, burrowssharp99, seagerwhitney00, seagersasselov00} that predicted the emergent spectra of hot Jupiters to be shaped by dominant radiative opacity from water vapor, carbon monoxide, carbon dioxide, and (for cooler and/or carbon-rich atmospheres) methane.  Figure~\ref{fig:189_spectrum} shows the agreement between the best available \emph{HST} \cite{crouzet14} and \emph{Spitzer} observations of HD\,189733b \cite{knutson09a,knutson12, kilpatrick19} and a cloud-free equilibrium chemistry model with parameterized pressure-temperature profile, solar atmospheric heavy element content (``metallicity"), and carbon to oxygen ratio \cite{zhang19}.  This planet was one of the earliest transiting hot Jupiters detected and is still one of the most favorable targets known today, and therefore provides a key testing ground for atmospheric models.

In its cryogenic phase, {\it Spitzer} had the capability to acquire low-resolution spectroscopy of hot Jupiters at secondary eclipse using the Infrared Spectrograph (IRS, see \citep{houck04}).  Due to {\it Spitzer's} modest aperture, IRS eclipse spectra were only possible for the two brightest systems, HD\,189733b \citep{grillmair07} and HD\,209458b \citep{richardson07, swain08}.  The initial results showed primarily continuous spectra, with little evidence for absorption features \citep{grillmair07, swain08}; however \citet{richardson07} found tentative evidence for silicate clouds in their spectrum of HD\,209458b near 9.65\,$\mu$m.  

Subsequent work on bright hot Jupiters has more completely defined their atmospheric chemistry. For HD\,209458b, {\it Spitzer} eclipse data in combination with ground-based cross-correlation spectroscopy \citep{brogi19} indicate a composition consistent with solar abundances, and with a carbon-to-oxygen ratio less than unity \citep{line16}.  The ultra-hot Jupiter WASP-12b was first observed during {\it Spitzer's} cryogenic phase \citep{madhusudhan11}, and seemed to have a carbon-to-oxygen ratio exceeding unity, i.e. it appeared to be carbon-rich ($C/O>1$).  However, \citet{line14} did not find $C/O > 1$ in a statistically convincing manner using a larger sample from {\it Spitzer} that included WASP-12b, and we discuss the $C/O$ issue in more depth in Section~\ref{subsec:trends} 

\subsection{A Search for Temperature Inversions}\label{subsec:temp}

Beyond the basic confirmation that hot Jupiter spectra were shaped by water vapor, carbon monoxide, carbon dioxide, and (in some cases) methane opacity, {\it Spitzer's} secondary eclipse data also provided constraints on the dayside pressure-temperature profiles of these atmospheres.  Many early {\it Spitzer} investigations reported evidence for the presence of ``stratospheres", otherwise known as temperature inversions, in hot Jupiter atmospheres \citep{madhusudhan10}. Although the default expectation is for temperature to decrease with increasing height, planets with a temperature inversion have a layer where temperature rises with increasing height.  In hot Jupiter atmospheres, these inversions were predicted to be caused by gas phase TiO or VO, which are strong optical absorbers \citep{burrows07, burrows08b, fortney08, spiegel09, madhusudhan10}.  However, subsequent photometry and spectroscopy of the archetypal inverted atmosphere (HD\,209458b, \citep{knutson08}) indicated that it did not in fact host a temperature inversion \citep{diamond-lowe14, line16}.  This change in interpretation was due to improved observing methods (early observations dithered the telescope, which increased the instrumental noise by nearly an order of magnitude) and better instrumental noise models (e.g., \citealp{deming15}). The current consensus based on a handful of planets observed with both \emph{Spitzer} and \emph{HST} is that temperature inversions due to TiO and VO {\it do} occur in hot Jupiter atmospheres, but only for the most highly irradiated ($>2000\,K$) planets \citep{kreidberg18, evans17}.

\subsection{Trends in Atmospheric Composition}\label{subsec:trends}

During {\it Spitzer's} initial cryogenic mission (Figure~\ref{fig:timeline}), it was possible to observe bright transiting planet systems in up to six photometric bands (e.g., Figure~\ref{fig:189_spectrum}); despite their low spectral resolution, these data sets nonetheless allowed for reasonably well-constrained inferences about atmospheric composition \citep[e.g.,][]{line16, madhusudhan11, morley17, stevenson14a}. {\it Spitzer} observed secondary eclipses for over 100 transiting planets in at least one wavelength band.  However, approximately 80\% of these planets were not observed until after the end of the cryogenic mission, when the telescope was limited to 3.6- and 4.5\,$\mu$m IRAC \citep{fazio04} photometry. With only two photometric points there are strong degeneracies between the atmospheric composition and pressure-temperature profile. For example, there can be multiple different combinations of atmospheric structure and composition that provide a comparably good match to the observed spectrum of a given planet. It is nonetheless possible to study the band-averaged brightness temperatures and spectral colors of this larger hot Jupiter ensemble statistically, using methods similar to the color-color diagrams used by stellar astronomers \citep{triaud14a, triaud14b, triaud15, beatty19, garhart20}. 

\begin{figure*}[h!]
\centering
\includegraphics[width=6in]{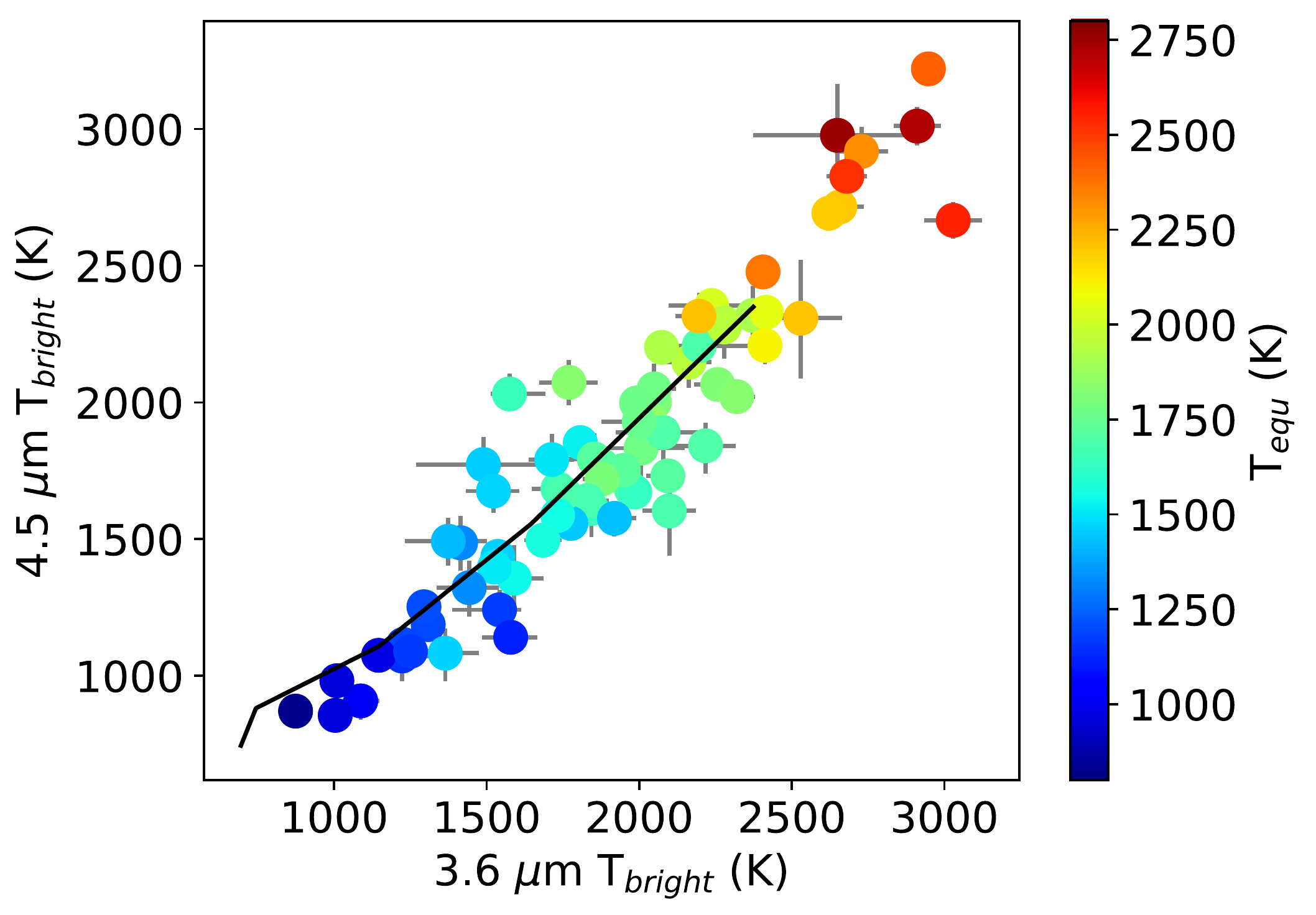}
\caption{Photometric brightness temperatures in the 3.6\,$\mu$m and 4.5\,$\mu$m \emph{Spitzer} bands from all published secondary eclipse observations of gas giant planets ($>2.5\sigma$ detection significance in both bands) versus predicted values from a representative grid of models \citep{fortney08} assuming zero albedo, full recirculation, and no dayside temperature inversions (i.e., excluding opacity from TiO and VO). Planet equilibrium temperature calculated assuming zero albedo and full recirculation is indicated by the color bar on the right.  Figure adapted by N. Wallack from \cite{garhart20} and Wallack et al. 2020, submitted.}
\label{fig:trend}
\end{figure*}

One of the strongest conclusions that can be drawn from the total of {\it Spitzer's} secondary eclipses involves the efficiency of longitudinal heat transport by winds.  {\it Spitzer} observations established that the most strongly irradiated planets circulate heat with the least efficiency, as we discuss in more detail in Section~\ref{subsec:phase_curves}.  In the realm of spectral shapes, \citet{garhart20} examined the ratio of the 4.5- to 3.6\,$\mu$m brightness temperature as a function of planetary equilibrium temperature for a sample of 37 hot Jupiters and found that the 4.5\,$\mu$m fluxes become more prominent relative to 3.6\,$\mu$m with increasing stellar irradiance (see Figure~\ref{fig:trend}). This runs counter to predictions from simple solar-composition equilibrium chemistry forward models (e.g., \citep{fortney06b}), suggesting that there may be systematic variations in hot Jupiter compositions or vertical thermal structures that are not captured by these models. 

Beginning with the Neptune-mass planet GJ\,436b (\citep{stevenson10, lanotte14, morley17}), a series of \emph{Spitzer} secondary eclipse studies \citep{kammer15,wallack19} have focused specifically on cooler ($T_{eq} < 1000$ K) transiting planet atmospheres, where methane is expected to replace carbon monoxide and carbon dioxide as the dominant carbon reservoir, resulting in a shift in the 3.6- to 4.5\,$\mu$m spectral slope.  At these temperatures, the ratio of atmospheric methane to carbon monoxide and carbon dioxide is predicted to be a sensitive function of atmospheric metallicity and carbon-to-oxygen ratio (e.g., \citep{moses13, drummond19}).  For GJ\,436b, whose dayside emission spectrum lacks any significant methane absorption and instead appears to have strong inferred CO and CO$_2$ absorption, the data are best-matched by models with a relatively high ($>200\times$ solar) atmospheric metallicity (e.g., \citep{stevenson10, moses13, morley17}). When compared to the ensemble of transiting planets with atmospheric metallicity constraints from \emph{HST} spectroscopy (e.g. \citep{kreidberg14, wakeford17, spake19}),  this planet appears to have one of the most metal-rich atmospheres observed to date.   

It has long been suggested that the atmospheric compositions of planets should reflect their formation locations and accretion histories.  In the solar system, the core mass fractions and atmospheric metallicities of gas giant planets are inversely correlated with their masses (e.g., \citep{lodders03}).  Mass and radius measurements for transiting gas giant planets indicate that small (i.e., Neptune-mass) planets also have a greater proportion of heavy elements in their bulk compositions than Jupiter-mass planets \citep{thorngren19}.  However, it is currently unclear whether or not planets with enhanced bulk metallicities also have enhanced atmospheric metallicities; some models predict that the answer may vary depending on the planet's migration history \citep{fortney13}.  \citet{wallack19} investigated trends in the 3.6- and 4.5\,$\mu$m brightness temperatures of transiting gas giant planets cooler than 1000\,K and found no evidence for a solar-system-like correlation between planet mass and atmospheric composition (e.g., \citep{kreidberg14}), but did identify a potential correlation (not statistically secure) between the inferred CH$_4$/(CO + CO$_2$) ratio and stellar metallicity.  These trends will be investigated in much greater detail by \emph{JWST}, which should provide a definitive answer to this question \citep{blumenthal18, bean18, schlawin18, drummond18}. 

\subsection{Carbon-to-Oxygen Ratio}\label{subsec:CtoO}

The ratio of carbon to oxygen in the atmospheres of transiting gas giant planets is also expected to vary with formation location.  Protoplanetary disk models predict that the carbon-to-oxygen ratio in the gas should vary with position in the disk, due to spatially different condensation of water and carbon monoxide \citep{oberg16, eistrup18}.  Hence $C/O > 1$ in an exoplanetary atmosphere is both plausible and can potentially provide useful constraints on a planet's formation location and accretion history.  The $C/O$ ratio of gas giant planet atmospheres - especially when it exceeds unity - profoundly alters the molecular composition of the exoplanetary atmosphere \citep{moses13, drummond19}, and can also impact the thermal structure \citep{molliere15} and cloud properties \citep{helling17}.  For atmospheres with $C/O > 1$, the water abundance is predicted to drop precipitously, with most of the oxygen being bound in formation of carbon monoxide. The absence of water vapor absorption, which is otherwise expected to dominate the observed spectra of these planets, should be obvious in transit and eclipse spectra, albeit less obvious in {\it Spitzer's} photometry. 

The first observational evidence for a high atmospheric $C/O$ ratio was reported by \citet{madhusudhan11}, who utilized {\it Spitzer's} secondary eclipse photometry to characterize the atmospheric composition of the very hot (2500\,K) Jupiter WASP-12b \citep{hebb09}.  \citet{stevenson14a} subsequently analyzed a more extensive set of {\it Spitzer} and {\it HST} secondary eclipse data for this planet and concluded that the evidence still favored a carbon-rich composition.  However, subsequent studies of this planet's transmission spectrum \citep{kreidberg15} indicated the presence of a strong water absorption feature, which appeared to run counter to this inferred high C/O ratio.  Nonetheless, more recent retrieval studies of the best available secondary eclipse data sets for this planet \citep{oreshenko17} continue to prefer models with relatively high carbon to oxygen ratios.  This tension between transmission and emission spectroscopy data serves to highlight some of the intrinsic degeneracies in retrievals based on low resolution spectroscopy and photometry.

\begin{figure*}[h!]
\centering
\includegraphics[width=6in]{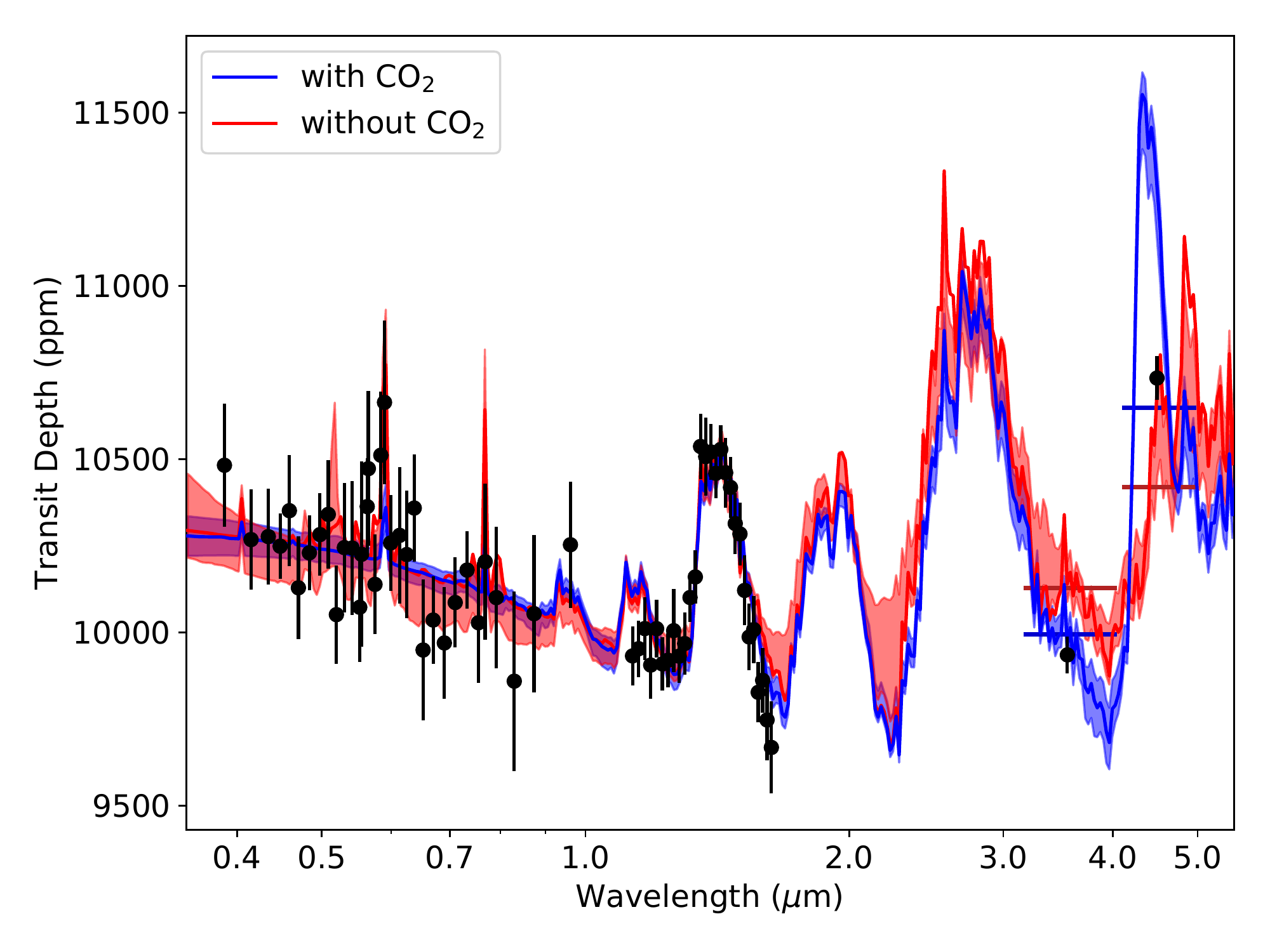}
\caption{Transmission spectrum for the sub-Saturn mass exoplanet WASP-127b.  \emph{HST} ($<2$ $\mu$m) and \emph{Spitzer} ($>2$ $\mu$m) data from \citet{spake19} are plotted as black filled circles, while best-fit open-source PLATON models \citep{zhang19}  with (blue) and without (red) CO$_2$ opacity are overplotted as solid lines, with uncertainties indicated as colored shading.  The band-integrated model values at 3.6 and 4.5\,$\mu$m are shown as horizontal dark red and blue lines. \citet{spake19} find that this planet has a super-solar atmospheric metallicity, and that the strong absorption (larger transit radius) in the 4.5\,$\mu$m band can only be matched when carbon dioxide is included in the models. Figure is courtesy of Y. Chachan.}
\label{fig:w127}
\end{figure*}

\subsection{Transit Spectroscopy}\label{subsec:transit_spectroscopy}

 Measurements of the wavelength-dependent transit depths, or ``transmission spectra" of transiting planets also provide complementary constraints on their atmospheric compositions.  Unlike secondary eclipses, which are strongly biased towards infrared wavelengths where the planet-star flux ratio is maximized, most transmission spectroscopy studies rely on observations spanning both optical and infrared wavelengths.  The overall amplitude of absorption features seen in transmission provides a constraint on the scale height of the atmosphere, which is a function of its mean molecular weight (e.g., metallicity). 
 
 Joint spectroscopy from {\it Spitzer} and Hubble (and also ground-based spectroscopy in many cases) helped to derive exoplanetary atmospheric metallicities for about a dozen exoplanets to date \citep{ehrenreich14, stevenson14b, nikolov15, fischer16, wakeford17, alam18, ducrot18, benneke19, sotzen19, spake19}.  Because the \emph{HST} coverage is limited to wavelengths where water is the dominant molecular absorber, {\it Spitzer} transit depths provide complementary constraints on absorption from methane, carbon monoxide, and carbon dioxide, all of which absorb strongly in the 3.6 and 4.5\,$\mu$m bands.  Recent observations of the sub-Saturn mass exoplanet WASP-127b shown in Figure~\ref{fig:w127}, from \citet{spake19}, illustrate the diagnostic power of combined \emph{HST} and \emph{Spitzer} data for constraining the abundances of carbon-bearing species.  In this case, {\it Spitzer's} transit radius at 4.5\,$\mu$m could only be matched with strong absorption by carbon dioxide.  \citet{benneke19} also leveraged \emph{Spitzer} transit data to show that methane was under-abundant relative to the predictions of equilibrium chemistry models in the atmosphere of the mini-Neptune GJ\,3470b. 

\subsection{Thermal Emission from Highly Irradiated Rocky Planets}\label{subsec:rocky}

Rocky exoplanets are much smaller than gas giants, and were therefore difficult targets for {\it Spitzer}.  However, a subset of these planets orbit extremely close to their host stars \citep{raymond14}, resulting in relatively high equilibrium temperatures.  Just a few years into its extended warm mission, \emph{Spitzer} became the first telescope to detect thermal emission from a super-Earth by combining multiple secondary eclipse observations of the ultra-short-period super-Earth 55 Cnc{\it\,e} \citep{demory12}. As of this writing, there is no unequivocal measurement of an atmosphere on an exoplanet that is definitively rocky, but there are intriguing hints.  One valuable technique that was pioneered by {\it Spitzer} investigators involves photometry of the exoplanet over its full orbit.  A so-called phase curve of a rocky exoplanet can in principle reveal the existence of an atmosphere, by demonstrating significant longitudinal heat transport \citep{seager09}.  Application of this method using {\it Spitzer} observations of the hot super-Earth 55 Cancri{\it\,e} indicated either an optically thick atmosphere or the existence of low-viscosity surface magma flows \citep{demory16b}.  In contrast, the same technique applied to the warm super-Earth LHS\,3844b indicated no atmosphere, or only a very thin atmosphere \citep{kreidberg19}.  

There is also some observational evidence to suggest that the dayside flux from at least one hot super-Earth (55 Cnc{\it\,e}) may vary significantly from orbit to orbit.  \citet{demory16a} observed a series of eight secondary eclipses of 55 Cancri{\it\,e} at 4.5\,$\mu$m and found that they varied by a large fraction of their average amplitude.  That conclusion was confirmed with an independent analysis of the same {\it Spitzer} data by \citet{tamburo18}.  Tamburo et al. suggest that the planet has a low albedo with inefficient heat redistribution intermittently covered over a large fraction of the substellar hemisphere by reflective grains, which could be produced by volcanic activity or variable clouds. 

\section{Atmospheric Dynamics}\label{sec:atm_dynamics}

{\it Spitzer} observers probed the dynamics of (primarily) hot Jupiter atmospheres using several observational techniques: phase curves, eclipse mapping, and searches for variability of secondary eclipse amplitudes.

\subsection{Thermal Phase Curves}\label{subsec:phase_curves}

\begin{figure*}[h!]
\centering
\includegraphics[width=6in]{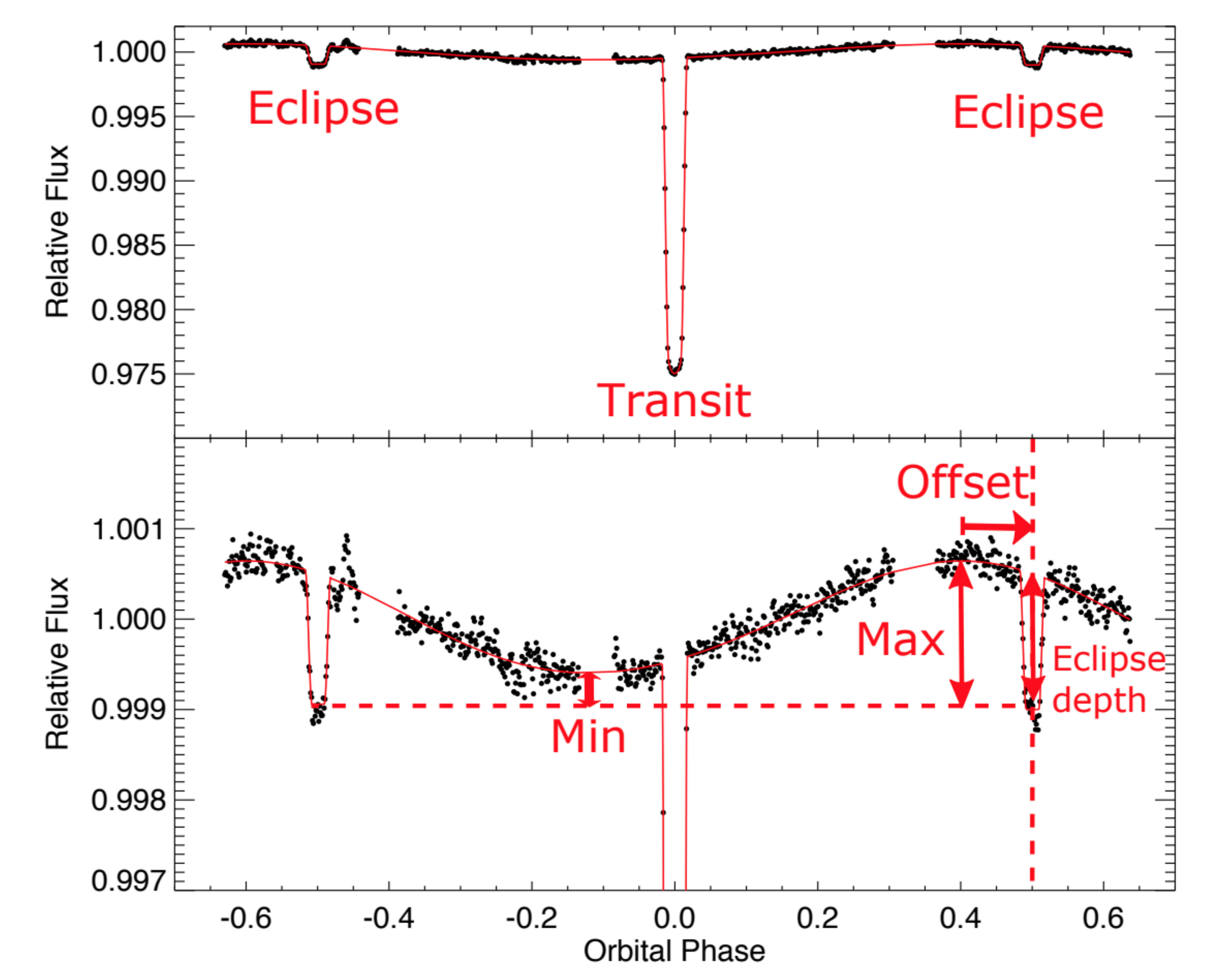}
\caption{Structure of an exoplanet phase curve, from \citet{parmentier18}.  These 3.6\,$\mu$m {\it Spitzer} data were acquired and analyzed by \citet{knutson12}.  }
\label{fig:phase_curve}
\end{figure*}

Arguably {\it Spitzer's} greatest impact on exoplanetary science came through the measurement of thermal phase curves.  For tidally locked planets, each orbital phase corresponds to a unique location on the planet, and the measured infrared brightness as a function of orbital phase can be inverted to produce a longitudinal brightness map for the planet (e.g., \citep{cowan08, knutson09a}, and many others cited below). Phase curve studies are also possible for non-transiting planets, but the information content of phase curves is greatest for observations of planets with known radii and orbital inclinations, and most studies have therefore focused on transiting planets. We show a representative 3.6\,$\mu$m \emph{Spitzer} phase curve for HD\,189733b \cite{knutson12} in Figure~\ref{fig:phase_curve} from \citep{parmentier18} in order to illustrate the relevant geometry.

The earliest phase curve measurements with \emph{Spitzer} began by combining a few discrete measurements spread over multiple epochs \citep{harrington06, cowan07}, but observers quickly realized that continuous phase curve monitoring allowed for both a higher signal-to-noise and more precise correction of instrumental noise sources.  The first full-orbit phase curve of the hot Jupiter HD\,189733b \citep{knutson07} spawned a flurry of additional phase curve observations, primarily of hot Jupiters \citep{knutson09a, knutson09c, knutson12, cowan12, lewis13, maxted13, lewis14, shporer14, wong14, zellem14, wong15, wong16, krick16, stevenson17, zhang18, dang18, mendonca18, kreidberg18, beatty19}.  For many of these planets, \emph{Spitzer} observed full-orbit phase curves in multiple bandpasses (typically just 3.6 and 4.5\,$\mu$m, but 8.0 and even 24\,$\mu$m phase curves exist for a few planets, including HD\,189733b \citep{knutson12}).  In principle, these multi-wavelength observations can be used to characterize the thermal emission spectra and corresponding atmospheric compositions, thermal structures, and cloud properties of these planets as a function of orbital phase \citep {drummond18, rauscher18, steinrueck19}.  However, in practice the limited number of bandpasses and relatively broad wavelength ranges of the \emph{Spitzer} photometric bands make atmospheric retrievals using phase curve data impractical. For a few planets, phase curves using both {\it Spitzer} and {\it HST} were analyzed to derive their dayside compositions (e.g., \citep{stevenson17, kreidberg18}).

In cloud-free atmospheric circulation models with equilibrium chemistry, both the amplitude of the phase curve and the offset of the peak are sensitive to atmospheric physics \citep{heng15}. In addition to numerical hydrodynamic models \citep{showman08, rauscher12, dobbs-dixon13, komacek16, komacek17, drummond18, tan19}, semi-analytic formulations have also been used to interpret the observations \citep{cowan11b}.  Planets with more efficient day-night circulation are expected to have larger phase offsets and smaller phase curve amplitudes, while those that are closer to radiative equilibrium will have little to no phase offset and relatively large phase curve amplitudes.  
General circulation models predict that more highly irradiated hot Jupiters should have less efficient heat transport than their more moderately-irradiated counterparts \citep{perez-becker13,komacek16}.  While it is true that most strongly irradiated planets have relatively large fractional phase curve amplitudes, there does not appear to be a tight correlation between phase curve amplitude and stellar irradiance \citep{parmentier18}.  This may indicate nightside clouds, which mask the signature of thermal emission in cloudy regions and increase the apparent day-night contrast in the \emph{Spitzer} bands \citep{keating19}.  In addition to transport by winds or waves, stellar energy in the most highly irradiated hot Jupiter atmospheres can be removed by dissociation of water vapor and molecular hydrogen on the hot day side \citep{parmentier18b, arcangeli18, lothringer18, tan19}.  Subsequent recombination can release that energy on the night side of the planet \citep{parmentier18, tan19}, thus augmenting hydrodynamic transport using chemistry. Observations of these ultra-hot Jupiters confirmed a lack of water absorption in their spectra, and revealed thermal inversions in several cases (WASP-18b, WASP-103b, HAT-P-7b \citep{sheppard17, arcangeli18, kreidberg18, mansfield18}).

Phase curve offsets appear to be more tightly correlated with irradiance level than are amplitudes, with the most highly irradiated planets showing relatively small phase offsets \citep{zhang18,parmentier18}. This means that the hottest portion of the day side atmosphere is located close to the substellar point for these planets, whereas in less-irradiated hot Jupiters this hot gas appears to be advected downwind (east) of the sub-stellar point by super-rotating equatorial winds, causing the phase curve to peak prior to the secondary eclipse.  This effect was first reported in \citep{knutson07}, whose phase curve observation of the hot Jupiter HD\,189733b provided observational confirmation for the existence of strong zonal winds on hot Jupiters \citep{showman08}.  The size of the offset is diagnostic of the radiative time scale compared to the time for transport of heat by wave motions or advection at the pressures probed by the \emph{Spitzer} bands \citep{perez-becker13, komacek17, parmentier18}.  Although most hot Jupiter phase curves have offsets to the east (i.e., super-rotating winds), CoRoT-2b has an offset to the west, possibly due to the presence of patchy clouds or magnetic effects \citep{dang18}. {\it Spitzer} phase curves can also be used to look for non-spherical planet shapes due to tidal effects and/or mass outflow.  \citet{bell19} used phase curve observations to detect ongoing mass loss on the ultra-hot archetype planet WASP-12b.   For this planet, the outflowing gas fills and emits within the planet's Roche lobe, whose solid angle changes as the planet orbits.

Although most hot Jupiters have closely circular orbits due to tidal circularization, a few are in eccentric orbits, with $e$ as great as  $0.52$ for HAT-P-2b \citep{lewis13, lewis14}, and $0.93$ for HD\,80606b \citep{fossey09}.  \citet{laughlin09} used {\it Spitzer} to measure the periastron passage of HD\,80606b.  They discovered that the planet also has a secondary eclipse (not necessarily true for eccentric transiting planets), and they made a quantitative measurement of the radiative time scale (also, see \citealp{dewit16}).  

\subsection{Trends in Atmospheric Circulation Efficiency from Secondary Eclipses}\label{subsec:dynamics_eclipses}

Secondary eclipses can also provide valuable insights into the longitudinal redistribution of heat on hot Jupiters.  Although \emph{Spitzer} observed full or partial phase curves for 26 exoplanets (not all are published yet), it observed secondary eclipses for more than 100 hot Jupiters, in many cases using at least two bandpasses.  Cowan and Agol \citep{cowan11} demonstrated that the brightness temperatures from these secondary eclipse depths can be used to infer statistical information about the nature of longitudinal heat redistribution and albedos of these planets.  These studies indicate that the most highly irradiated hot Jupiters have relatively high dayside brightness temperatures, requiring both low albedos and inefficient day-night circulation, while less irradiated planets appear to have more efficient circulation and/or higher albedos \citep{schwartz15, schwartz17, garhart20}.  These observations are in good agreement with results from general circulation models, which predict a trend of decreasing circulation efficiency with increasing irradiation \citep{perez-becker13, komacek16}.

\subsection{Eclipse Mapping and Variability}\label{subsec:mapping}

At very high signal-to-noise, secondary eclipse observations can also be used to directly map the dayside brightness distributions of transiting planets.  This was first pointed out in pioneering work by \citet{williams06}, who noted that a non-uniform star-facing hemisphere will cause an apparent time lag on the order of tens of seconds between the observed secondary eclipse phase and the phase predicted for a spatially uniform planet.   This time lag was first detected observationally for the hot Jupiter HD\,189733b \citep{agol10} with a direction and magnitude consistent with phase curve results for that planet \citep{knutson12}.  Taking this phenomenon one step further, the variation in flux as the planet is gradually occulted can be inverted to yield a spatial map of the dayside (star-facing) hemisphere of the planet \citep{dewit12, majeau12} (not to be confused with phase curves maps covering all longitudes).  The hot Jupiter HD\,189733b is the only exoplanet whose dayside atmosphere was mapped in this fashion.  These initial results \citep{dewit12, majeau12} show an eastward hot spot, consistent with results from phase curve observations.  Subsequent improvements in this mapping technique \citep{rauscher18} confirmed this basic result. 

In the temporal domain, observers monitored secondary eclipses of the two brightest transiting hot Jupiter systems (HD\,189733b and HD\,209458b) to search for possible temporal variability \citep{agol10, kilpatrick19}.  The observed upper limits ($\lessapprox 6\%$) are several times greater than predictions from hydrodynamic models \citep{komacek19}.

\section{Properties of Orbits}\label{sec:orb_dynamics}

\subsection{Eccentricities}\label{subsec:eccentricities}

{\it Spitzer's} precise transit and secondary eclipse observations can also be used to probe the orbital properties of exoplanetary systems. It has been suggested that hot Jupiters may have formed at much larger orbital separations and then migrated inward via disk integration or high eccentricity migration and circularization \citep{dawson18}. Because tidal circularization is predicted to be slow, the frequency of residual non-zero eccentricities for hot Jupiters (as a function of semi-major axis) can in principle constrain the likelihood of a high eccentricity migration channel \citep{dawson18}. However, there are multiple ways for a planet to acquire a non-zero orbital eccentricity (planet-planet scattering, secular dynamics, disk interactions, etc.). To distinguish between the signatures of various mechanisms, sensitivity to small orbital eccentricities ($e \sim 0.01$) is desirable, but difficult to achieve using radial velocity observations alone.  Fortunately, secondary eclipse timing observations from {\it Spitzer} (in combination with radial velocities) yielded precise eccentricity estimates \citep{deming07, blecic13, lewis13, knutson14}, and limits on eccentricity \citep{knutson09b, todorov10, deming11} for hot Jupiters.  {\it Spitzer's} secondary eclipse times often give $e\cos{\omega}$ to a precision better than $0.01$, but not $e$ directly.  However, the argument of periastron ($\omega$) should be distributed randomly, hence secondary eclipse times are statistically useful to define the residual eccentricity distribution as a function of semi-major axis.  Those statistical studies are just beginning \citep{garhart20}, but there are ample {\it Spitzer} eclipse data that can be utilized, especially when orbital ephemerides can be improved using TESS transits.

Beyond statistical studies, {\it Spitzer} was instrumental in probing the properties of individual planets.  We here highlight two examples: the interior structure of HAT-P-13b, and the orbital decay of WASP-12b.

\subsection{The Core Mass of HAT-P-13b}\label{subsec:hat13}

The HAT-P-13 system comprises a hot Jupiter (HAT-P-13b) and two companion planets with much longer orbital periods, one of which has an eccentric orbit \citep{winn10b}.  That particular orbital configuration will produce a slight eccentricity in the orbit of HAT-P-13b, and the magnitude of that eccentricity depends on its internal mass distribution, specifically on the core mass \citep{batygin09}.  This is a key question for formation models, but while it is possible to constrain the bulk metallicities of hot Jupiters using masses and radii \citep{thorngren19}, these observations do not provide information about the relative distribution of these metals between the core and envelope.  Radial velocity observations \citep{winn10b} indicated a small non-zero eccentricity for HAT-P-13b. Secondary eclipse times using {\it Spitzer} confirm a small eccentricity, but the core mass is sensitive to the exact phase of the eclipse. \citet{buhler16} find a probable core mass of about 11 Earth masses, whereas \citet{hardy17} concluded that the core is small or non-existent, but they note that their eclipse times are significantly inconsistent (differing by 23 minutes) between the two {\it Spitzer} bandpasses.  Independently measured eclipse times from \citet{garhart20} are internally consistent and agree with Buhler et al., thereby supporting the 11 Earth mass core estimate.  

\subsection{The Orbital Decay of WASP-12\,b}\label{subsec:wasp12_orbit}

\begin{figure*}[h]
\centering
\includegraphics[width=6in]{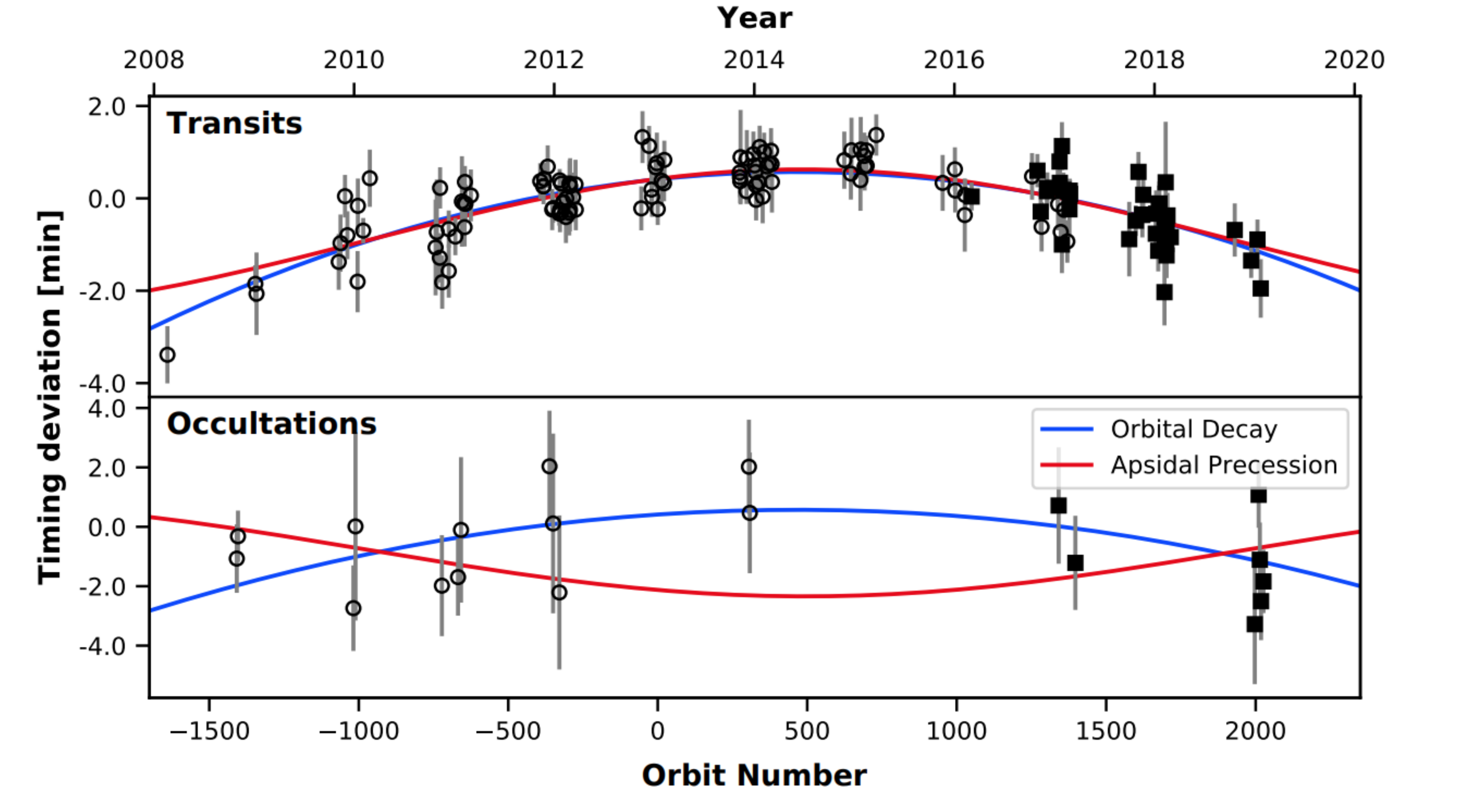}
\caption{Transit and secondary eclipse (occultation) times for WASP-12b, showing orbital decay, from \citet{yee19}.  The upper panel shows residuals of transit times from a linear ephemeris; the curvature shows that the period derivative is not zero.  The lower panel shows timing residuals for primarily {\it Spitzer's} secondary eclipses, and models for orbital decay versus apsidal precession.  Orbital decay is highly favored over apsidal precession. }
\label{fig:wasp12}
\end{figure*}

The orbits of hot Jupiters should be decaying as tidal dissipation removes energy from their orbits.  For hot Jupiters with the shortest known periods, the orbital decay is astrophysically fast, but long on a human time scale.  Nevertheless, {\it Spitzer} made orbital decay possible to observe in the case of WASP-12b, a very close-in and ultra-hot Jupiter.  \citet{patra17} found that the orbital period was apparently decreasing by $29\pm3$ milli-seconds per year, based on ground-based transits and {\it Spitzer} secondary eclipses.  However, they could not rule out apsidal precession, wherein the orientation of the orbit within the orbital plane changes, but the orbital period remains constant.  Fortunately, secondary eclipses could distinguish these possibilities, and {\it Spitzer} secondary eclipses \citep{yee19} confirmed the decrease of the orbital period, as shown in Figure~\ref{fig:wasp12}.  The results give insight into the physics of tidal dissipation, specifically the Q-factor \citep{goldreich66} of the star (the dissipation occurs within the star).  \citet{yee19} find a Q-factor of $1.75 \times 10^5$, which is lower than many previous (but less direct) determinations.  Yee et al. also use new radial velocity data to prove that the observed acceleration is not due to changes in light travel time caused by a companion planet in a long period orbit. 

\subsection{Systems of Planets}\label{subsec:systems}

\begin{figure*}[h]
\centering
\includegraphics[width=6in]{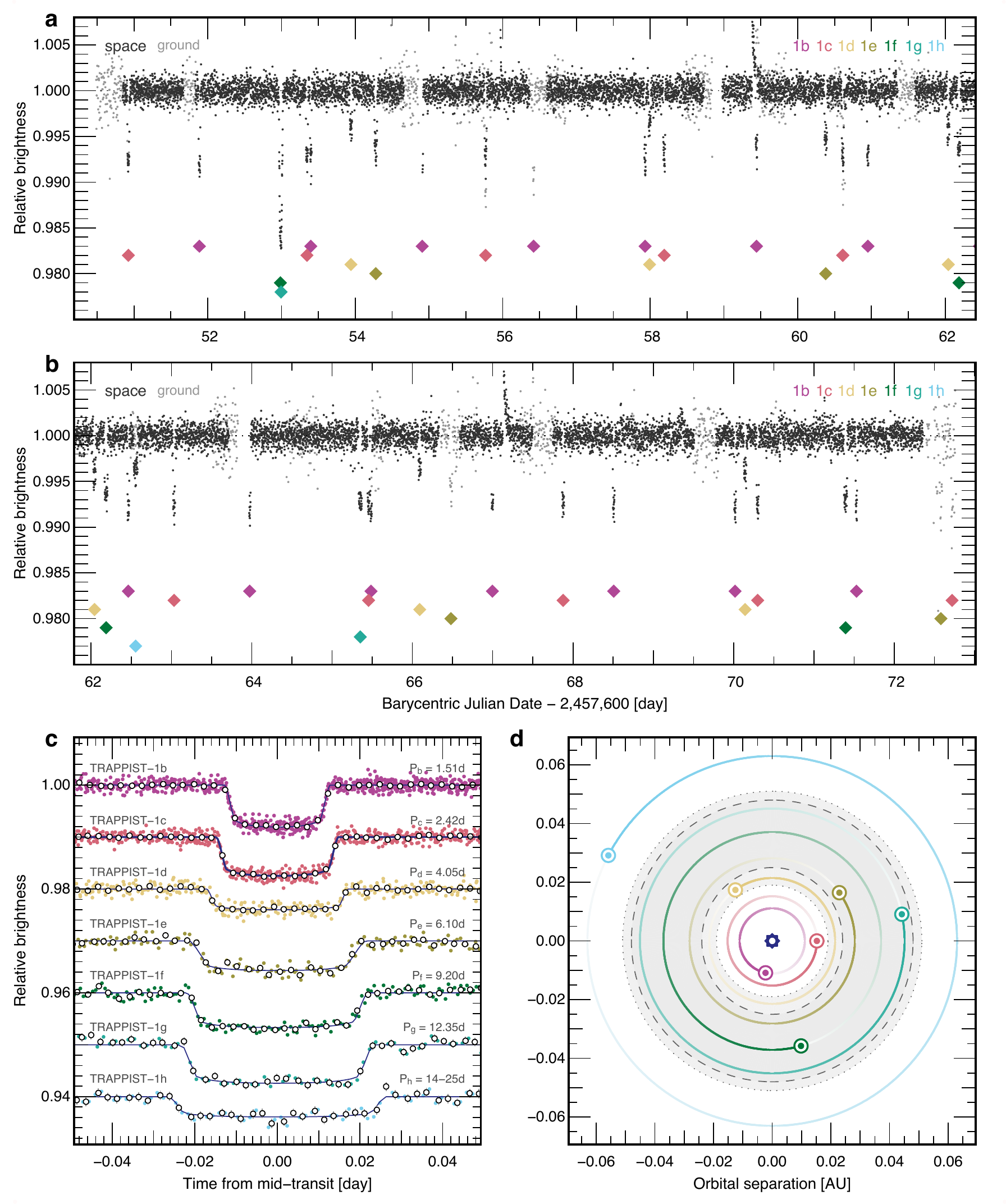}
\caption{Transits of planets in the TRAPPIST-1 system as observed with {\it Spitzer} and ground-based photometry, from \citet{gillon17c}.  The upper two panels (a and b) show the photometry, with colored symbols marking transits of the various planets.  The lower left panel (c) shows the phased transits of all seven planets, and the lower right panel (d) shows their orbits. }
\label{fig:trappist}
\end{figure*}

{\it Spitzer} was especially valuable in searching for transits of planets orbiting M-dwarf stars, because M-dwarfs are bright in the IR.  The M-dwarf star GJ\,1214 hosts a transiting mini-Neptune \citep{charbonneau09}, orbiting closer to the star than the nominal habitable zone (HZ).  \citet{fraine13} and \citet{gillon14} searched for planets in the inner HZ of GJ\,1214, and placed Mars-sized upper limits on the presence of such planets.  However, the largest payoff for {\it Spitzer} was the delineation of multiple transiting planets orbiting the ultra-cool M-dwarf system TRAPPIST-1.  Discovered by the ground-based TRAPPIST survey \citep{gillon16}, a long-duration quasi-continuous sequence of {\it Spitzer} photometry \citep{gillon17c} revealed a system of 7 rocky planets, all transiting the small M-dwarf star (Figure~\ref{fig:trappist}).   Exoplanetary scientists have already begun characterizing the atmospheres of these worlds \citep{dewit18, durcot18}, but no unequivocal atmospheric detections have yet been achieved. Their atmospheric transmission spectra are predicted to (potentially) contain absorption features from molecular oxygen, ozone, water vapor, sulphur dioxide, carbon monoxide, and methane \citep{lincowski18, hu20}, and it is possible that the planets may be tidally heated \citep{dobos19}, or heated inductively via the stellar magnetic field \citep{kislyakova17}.  There is enormous community interest in the TRAPPIST-1 planets, and they are expected to be important targets for JWST \citep{morley17b, lustig-yaeger19}.

\subsection{Transit Timing Variations in the TRAPPIST-1 System}\label{subsec:ttvs}

When multiple planets are present in a system, their mutual gravitational perturbations produce transit timing variations (TTVs), manifest as departures from a strictly linear ephemeris \citep{holman05}.  Those TTVs can be used, in conjunction with a dynamical model, to infer the masses of the planets \citep{agol18}. In many cases, TTVs are the only practical method to derive exoplanet masses because small planets often produce a radial velocity signal in the stellar spectrum that is too small to measure. In contrast, TTVs can be readily measured, especially using long continuous photometric sequences from {\it Spitzer}. In the case of multi-planet systems such as TRAPPIST-1, mutual occultations among the planets can help to extract precise TTVS \citep{luger17}.  \citet{delrez18} and \citet{grimm18} analyzed 60 and 284 transit times, respectively, for the TRAPPIST-1 system, and Grimm et al. derived masses with precisions between 5\% and 12\%.  They used those masses to infer that two of the planets were predominately rocky, while the remaining five have low density envelopes such as atmospheres, oceans, or layers of ice (also, see \citealp{dorn18}).  Those inferences are valuable in planning atmospheric characterization studies of the system using JWST.

\subsection{Planetary Radii and Orbital Periods}\label{subsec:radii}

The first benefit of a primary transit is to obtain a precise radius for the transiting exoplanet, and accurate radii are fundamental for characterizing exoplanetary properties.  Stellar limb darkening is greatly reduced in the IR compared to optical wavelengths. Consequently, {\it Spitzer's} transits tend toward simple box-like shapes, and yield the ratio of planet-to-star radius in a simple and minimally model-dependent manner, albeit with potentially higher random noise due to reduced stellar photon fluxes in the IR \citep{richardson06, nutzman09, gillon12}.  The high cadence and uninterrupted photometry available from Spitzer were crucial for precise transit measurements \citep{demory11, ballard14, chen18}, especially when the transit had a long duration \citep{hebrard10}.  Moreover, star spots and plage - which can potentially interfere with accurate radius measurements - have low thermal contrast with stellar photospheres in the IR \citep{fraine14, morris18}, further increasing the utility of {\it Spitzer's} measurements of exoplanetary radii.  Although solar-type stars are not as bright in the IR as in the optical, M-dwarf stars provide high fluxes in {\it Spitzer's} bands, allowing precise radii for their (sometimes small) transiting planets \citep{gillon07, fraine13, gillon16, chen18}. 

After a transiting planet is discovered, imprecision in its orbital period leads to an accumulating error in the times of transit and eclipse as time passes.  In order to observe transiting planets with JWST, it is necessary to have accurate orbital periods, and {\it Spitzer} played a key role in that effort.  Follow-up of Kepler and K2 planets, for example, was possible with Spitzer, improving the orbital periods and in some cases measuring TTVs \citep{beichman16, benneke17, berardo19, dalba19, livingston19}. 

\section{Other Techniques}\label{sec:other}
Although the majority of exoplanets studied by Spitzer were in transiting systems, the observatory also enabled significant work using other techniques such as high contrast imaging and microlensing.

\subsection{High Contrast Imaging}\label{subsec:imaging}
{\it Spitzer's} modest aperture provided relatively low spatial resolution: the diffraction-limited full-width-to-half-maximum of point sources is 1.3 seconds of arc at 4.5\,$\mu$m wavelength.  Nevertheless, {\it Spitzer} had excellent sensitivity to low flux levels at thermal wavelengths where young, hot planets will emit.  That motivated searches for giant planets at large orbital distances.  Those searches included specific bright stars such as Vega, Fomalhaut, and Epsilon Eridani \citep{janson15}, as well as larger samples of young stars, including many host stars of known exoplanets \citep{durkan16}.  Although {\it Spitzer} did not detect any new exoplanets by imaging, the surveys provided important constraints on planet formation at distances between 100 and 1000 AU \citep{durkan16}.

In addition to exoplanet imaging searches, observers also used {\it Spitzer} to discover and investigate especially interesting companions in the brown dwarf mass range.  \citet{leggett10} obtained IRS spectroscopy of a T8 brown dwarf in a binary system with an M-dwarf star.  They used this spectrum to place constraints on the surface gravity, inferring a mass between $24-45$ Jupiter masses.  \citet{luhman11, luhman12} found a very cool companion to the white dwarf WD\,0806-661 with a temperature of 300 K and a corresponding mass of approximately 7 Jupiter masses.

\begin{figure*}[h]
\centering
\includegraphics[width=6in]{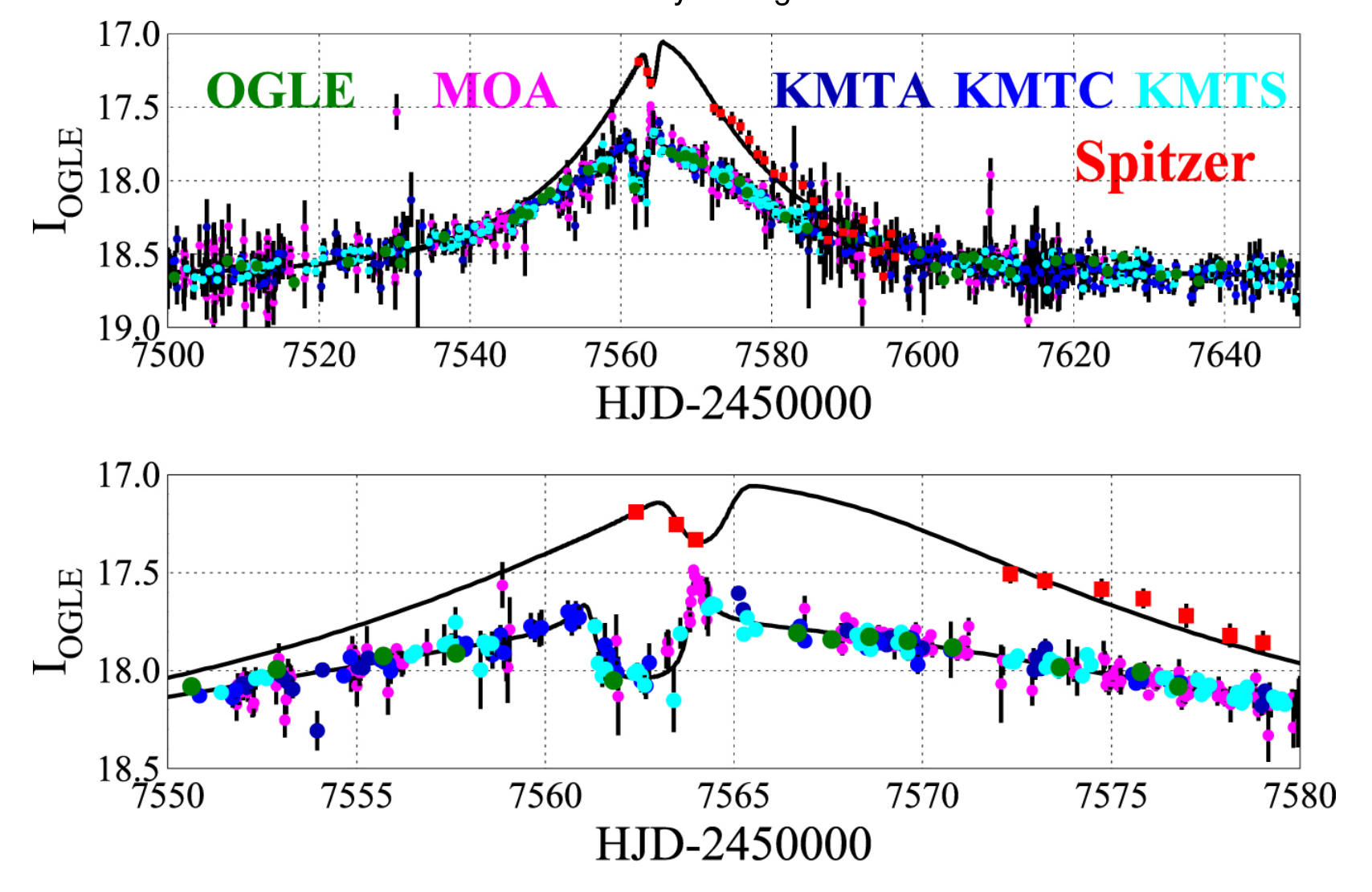}
\caption{Microlensing light curves for OGLE-2016-BLG-1067, from \citet{calchinovati19}.  This is a 0.43 Jupiter mass planet orbiting an M-dwarf star, at a projected distance of 1.7 AU (beyond the snow line in the M-dwarf system).  Note the very different light curve from {\it Spitzer} (red points), compared to the ground-based observations. }
\label{fig:microlensing}
\end{figure*}

\subsection{Microlensing}\label{subsec:microlensing}
Even prior to the launch of {\it Spitzer}, it was anticipated to be an important facility for microlensing \citep{gould99}.  {\it Spitzer's} continuous and nearly uninterrupted viewing enable photometry that can define the structure of microlensing events, and thereby determine the nature of the lensing systems, including the presence of planets.  Microlensing has a unique ability to detect small planets at large orbital distances, inaccessible to other techniques \citep{gaudi12}. Accurate photometry is difficult in the crowded fields that are used for microlensing searches.  Nevertheless, the microlensing studies were productive in finding planets in parts of the galactic disk \citep{street16} and bulge \citep{ryu18} at distances of several kiloparsecs.  Moreover, {\it Spitzer's} drift-away orbit gives it a view of microlensing events over a long spatial baseline when combined with observations from the ground \citep{shvartzvald17}, or from another spacecraft \citep{shvartzvald16}.  An example of the photometrically different view of a microlensing light curve by {\it Spitzer} is illustrated in Figure~\ref{fig:microlensing}, from \citet{calchinovati19}.   The different view from {\it Spitzer} permits the measurement of the microlensing parallax \citep{gould99, udalski15}, from which the masses of the lensing star and planet (not merely their ratio), and their projected orbital separation, can be determined.  

Results from {\it Spitzer's} microlensing campaigns include both giant planets \citep{calchinovati19} as well as low mass planets approaching the mass of Earth \citep{gould19}. \citet{shvartzvald17} point out that, together with the TRAPPIST-1 system, their microlensing detection of a 1.4 Earth-mass planet orbiting an ultra-cool M-dwarf star suggests that systems of rocky planets may be common around ultra-cool M-dwarf stars. The ultimate goal of the microlensing studies is to understand the frequency of occurrence of exoplanets in different regions of the Galaxy, and {\it Spitzer} made significant advances toward that goal \citep{dang19}.

\section{Brown Dwarfs} \label{sec:brown_dwarfs}

Beyond exoplanets, {\it Spitzer} also advanced the study of brown dwarfs.  These substellar objects can overlap with giant exoplanets in mass, but even when found in orbit around a more massive star they are believed to be a separate population. Most planets and brown dwarfs form by different mechanisms, but massive planets are sometimes found at orbital distances of 100's to 1000's of AU \citep{nielsen19, bowler20}, and those planets may form in a similar manner as brown dwarfs \citep{kratter16, kouwenhoven20}.

{\it Spitzer's} brown dwarf investigations can be broadly divided into several sub-topics. First, {\it Spitzer} measured parallaxes for brown dwarfs, which made it possible to determine fundamental properties such as mass.  Second, {\it Spitzer} mapped the nature of their emergent spectra, helping to extend the stellar spectral sequence to lower mass objects (L, T,  and Y spectral classes). {\it Spitzer} observers also mapped weather on brown dwarfs via their rotational light curves.  We discuss these topics in more detail below.  Moreover, {\it Spitzer} imaging was used to discover new brown dwarfs, often as companion to brighter stars, and those cases are discussed in Section~\ref{subsec:imaging}.  

\subsection{Parallaxes and Proper Motions}

When studying brown dwarfs, measurements of common proper motion in binary systems and distances via parallax are often crucial.  Observers exploited the long time baseline and varying orbital position of {\it Spitzer} to make both common proper motion \citep{luhman12} and parallax measurements \citep{kirkpatrick13, martin18, kirkpatrick19} for brown dwarfs, sometimes in combination with HST and/or the WISE mission \citep{beichman14}. {\it Spitzer's} parallax measurements for Y-dwarfs are especially important \citep{martin18, kirkpatrick19}, because distances are crucial for inferring properties such as mass. For example, \citet{leggett17} determined that the coldest known Y-dwarf (WISE 0855-0714, $\sim$\,250 K) has a mass between $1.5-8$ Jupiter masses, based on a {\it Spitzer} parallax that enabled comparison with evolutionary models. Spitzer parallaxes \citep{kirkpatrick19} were critical to determining the mass function of brown dwarfs, and showing that the low mass cutoff for their formation is probably less than 5 Jupiter masses.  This indicates overlap between the core accretion and disk fragmentation populations in a low mass range. 

\subsection{Emergent Spectra of Brown Dwarfs}

Brown dwarfs give us the opportunity to study and model the spectra of Jovian-mass objects without the complexity of starlight rejection by either coronagraphy or transits, or the atmospheric reaction to strong stellar irradiation. They thereby offer a laboratory for atmospheric modeling and comparison to models used for core accretion systems. Brown dwarf spectroscopy is thereby expected to be an important topic for JWST \citep{morley19}.

In {\it Spitzer's} cryogenic phase, \citet{roellig04} used IRS to obtain spectra of an M, L, and T dwarf (one in each class).  Although water vapor absorption was present in those spectra, the highlight was absorption by methane in the band at 7.8\,$\mu$m, and also the first detection of ammonia absorption (near 10.5\,$\mu$m).  Subsequent studies using larger samples of brown dwarfs \citep{cushing06, leggett09} showed that absorption by methane and ammonia appears at the L/T transition, with signatures of silicate and iron condensate clouds also being common for L and T dwarfs.  \citet{saumon06} studied ammonia absorption in a T7.5 dwarf, and could only account for the spectrum by reducing the ammonia abundance approximately one order of magnitude below a chemical equilibrium model.  They attributed their result to disequilibrium caused by vertical mixing.   Extending brown dwarf studies to the Y dwarfs, \citet{leggett17} also concluded that vertical mixing is important, and they derived effective temperatures, surface gravities, and metallicities for four Y-dwarfs with temperatures close to 600\,K.  

In addition to IRS spectra, photometry using IRAC \citep{patten06, burningham13, leggett17} defined the position of brown dwarfs in color-magnitude and color-color diagrams. Although studying brown dwarf colors was not new {\it per se}, {\it Spitzer's} sensitive observations in new wavelength bands produced a new perspective on the colors of brown dwarfs. These studies show that not only temperature, but also mass (via surface gravity) and metallicity affect brown dwarf colors, and they again found that departures from chemical equilibrium are important \citep{leggett17}.   Comparing brown dwarfs to hot Jupiters, \citet{beatty14} found that isolated brown dwarfs have colors that are very similar to the hot Jupiters and to the irradiated brown dwarf KELT-1b.

Brown dwarfs can be highly variable in their thermal emission. \citet{esplin16} used IRAC photometry to find variability in the IR emission of WISE 0855-0714. \citet{morales-calderon06} used IRAC photometry in a pioneering search for photometric variability due to weather patterns on rotating brown dwarfs.  Variability of brown dwarfs was subsequently detected in ground-based observations \citep{artigau09}.  Investigations using {\it Spitzer} in combination with {\it HST} and/or ground-based photometry exploited different heights of formation versus wavelength for powerful probes of inhomogenous cloud patterns as brown dwarfs rotate \citep{buenzli12, apai13, yang16, leggett16, biller18}.  {\it Spitzer's} capability for long uninterrupted observational sequences was key to studying variability of brown dwarfs. The magnitude of variability is a function of the viewing angle (pole versus equator \citep{vos17}), and the largest variations are seen when the line of sight is near-equatorial.  For example, \citet{biller18} observed the full amplitude of variability in the young planetary-mass object PSO J318.5-22 using IRAC 4.5\,$\mu$m in combination with {\it HST} at 1.1-1.7\,$\mu$m.  They found a large phase offset between the {\it Spitzer} and {\it HST} wavelengths, attributed to different longitudinal cloud structures at different pressures (each layer has a distinct temperature and corresponding infrared emission).  The rotational variations can be complex, and the patterns can vary on time scales longer than the rotational period \citep{apai17}.

\begin{figure*}[h]
\centering
\includegraphics[width=6in]{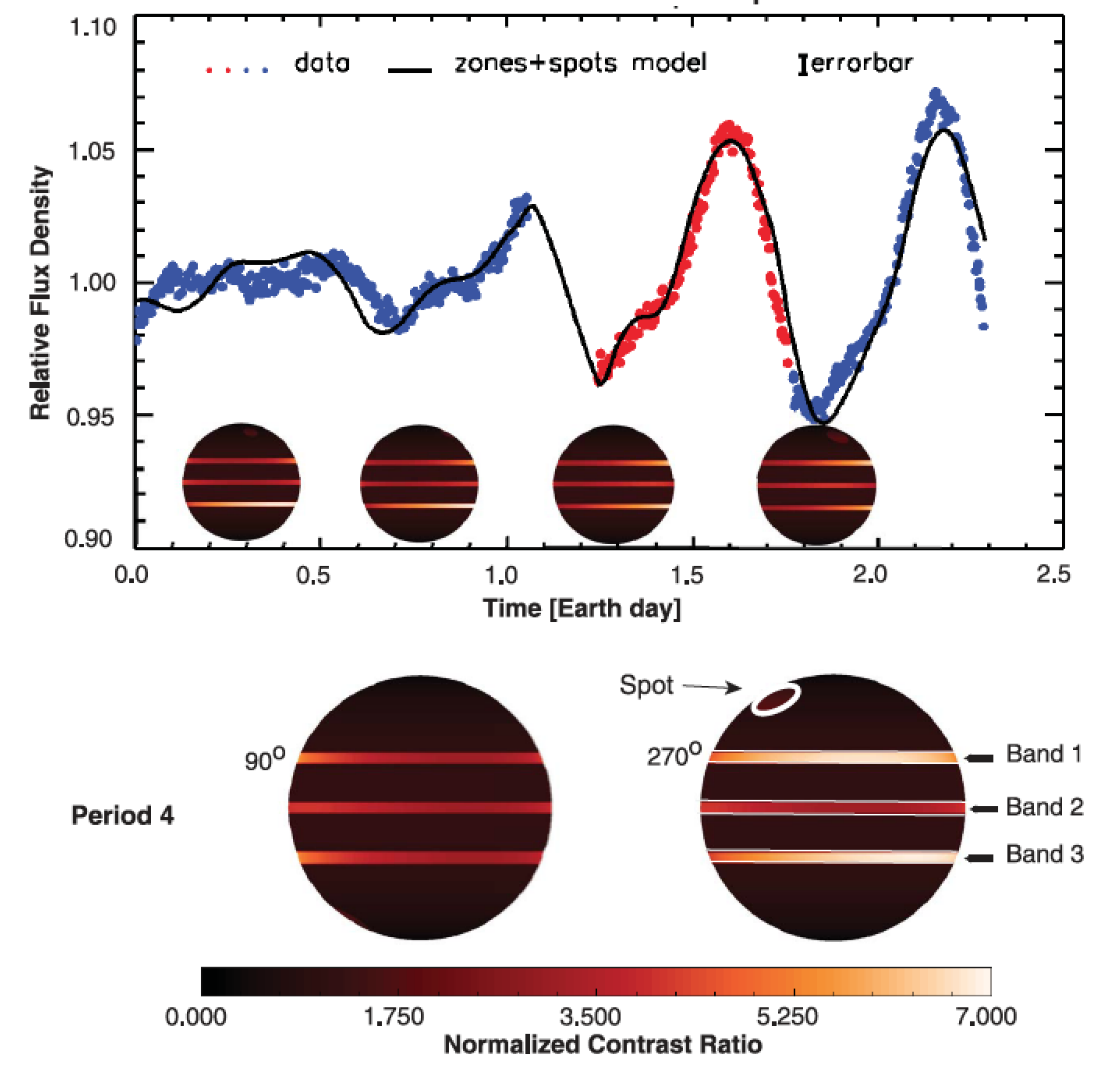}
\caption{ {\it Spitzer} IRAC photometry of 2MASS J21392216+0220185, from \citet{apai17}.  The top panel shows the photometry (blue points = 3.6\,$\mu$m, red points = 4.5\,$\mu$m), and the lower panel shows the retrieved model having three bands and a spot.  Due to zonal winds, the bands have different rotation periods, and beat against each other in the integrated light.}
\label{fig:brown dwarf_rotation}
\end{figure*}

\citet{apai17} analyzed long term IRAC photometry to infer the presence of planetary-scale wave features and discrete spots on 2MASS J21392216+0220185, illustrated on Figure~\ref{fig:brown dwarf_rotation}.  They infer the presence of bands whose brightness varies longitudinally (e.g., from variations in cloud opacity).  The rotational period differs slightly from band to band, due to zonal winds.  Those different periods cause a beating effect that is revealed in the {\it Spitzer} photometry.   {\it Spitzer} observers have thereby demonstrated a rich dynamic meteorology in brown dwarf atmospheres. The {\it Spitzer} results stimulate interest in continuous spectral monitoring with high sensitivity using JWST. Finally, we note that brown dwarf meteorology is similar to variable bands seen on Neptune and Jupiter \citep{apai17}, illustrating a link between low mass brown dwarfs and planets. 

\section{The Stage is Set for JWST} \label{sec:stage}

\subsection{Operational Legacy}\label{subsec:operations}

The Spitzer Science Center (SSC), in partnership with the broader exoplanet community, pioneered many innovative operational techniques to maximize the scientific yield of the mission for exoplanetary science.  Those innovations include improvements in the duration of observational sequences and in data compression \citep{carey11}, mitigation of pointing fluctuations \citep{carey11, ingalls12, carey14, grillmair14}, and the development of novel techniques to remove instrumental noise \citep{ingalls18}.  Frequent interactions between observers and the SSC culminated in a ``data challenge" to test a multitude of methods for precise and accurate correction of instrumental noise sources in IRAC photometry \citep{ingalls16}.  Those efforts influenced plans for JWST's Early Release Science program \citep{bean18}, and we expect that the operational lessons from {\it Spitzer} will be an enduring legacy for JWST.

\subsection{Scientific Legacy}\label{subsec:scientific} 

 Without observations from {\it Spitzer}, the potential for exoplanetary science from JWST would be far less clear.  {\it Spitzer} defined the magnitude of infrared emission from hot Jupiters, constrained the nature of their emergent spectra, and sharpened the questions concerning their atmospheric physics and chemistry.  {\it Spitzer's} sensitivity to carbon-bearing molecules was a prelude to the new insights that will be possible from JWST's panchromatic spectra.  The spectra of hot Jupiters change in response to stellar irradiation in ways that we currently do not fully understand, and their phase curves exhibit an interplay between radiative heating, cooling, and cloud formation that challenges our hydrodynamic models.   Beyond hot Jupiters, {\it Spitzer} probed Neptunes and super-Earths, finding new phenomena such as disequilibrium chemistry and puzzling variability in day side emissions. In the study of brown dwarfs, {\it Spitzer} measured weather patterns via rotational variability.  {\it Spitzer} also discovered new ultra-cool brown dwarfs and characterized their emergent spectra, distances, and space densities.   
 
Observations from {\it Spitzer} also probed the orbital dynamics of close-in exoplanets, with implications for the internal structure of the planets and their host stars.  {\it Spitzer} mapped planetary systems such as TRAPPIST-1, and measured masses using TTVs.  {\it Spitzer's} observations opened multiple sub-disciplines of exoplanetary science, to a degree not dreamed of before its launch.  JWST will begin with a rich menu of fascinating questions that are the legacy of exoplanetary science from {\it Spitzer}.
\\
\\
{\bf Acknowledgements.} We thank an anonymous referee for a critique that improved this paper. We are grateful to Dr. Mark Marley for his comments on the brown dwarf section, and Dr. Eric Agol for his comments on the TRAPPIST-1 masses.  We thank Yayaati Chachan, Nicole Wallack, and Michael Zhang for making Figures.
\\
\\
{\bf Author contributions.}  Both authors worked on writing the text, and selecting the figures and references.
\\
\\
{\bf Competing interests.}  The authors declare no competing financial interests.

\clearpage
\bibliography{references}

\end{document}